# Discovery of Oxygen Kα X-ray Emission from the Rings of Saturn

Anil Bhardwaj[1,*], Ronald F. Elsner[1], J. Hunter Waite, Jr.[2], G. Randall Gladstone[3], Thomas E. Cravens[4], and Peter G. Ford[5]

[1] NASA Marshall Space Flight Center, NSSTC/XD12, 320 Sparkman Drive, Huntsville, AL 35805, USA; anil.bhardwaj@msfc.nasa.gov, ron.elsner@msfc.nasa.gov

[2] Department of Atmospheric, Oceanic and Space Sciences, University of Michigan, Ann Arbor, MI 48109, USA; hunterw@umich.edu

[3] Southwest Research Institute, San Antonio, P.O. Drawer 28510, TX 78228, USA; randy.gladstone@swri.org

[4] Department of Physics and Astronomy, University of Kansas, Lawrence, KS 66045, USA; cravens@ku.edu

[5] Massachusetts Institute of Technology, Kavli Institute for Astrophysics and Space Research, 70 Vassar Street, Cambridge, MA 02139, USA; pgf@space.mit.edu

[*] on leave from: Space Physics Laboratory, Vikram Sarabhai Space Centre, Trivandrum 695022, India; anil_bhardwaj@vssc.org

Running header: X-ray emission from Saturn's rings

**Abstract**

Using the Advanced CCD Imaging Spectrometer (ACIS), the Chandra X-ray Observatory (CXO) observed the Saturnian system for one rotation of the planet (~37 ks) on 20 January, 2004, and again on 26–27 January, 2004. In this letter we report the detection of X-ray emission from the rings of Saturn. The X-ray spectrum from the rings is dominated by emission in a narrow (~130 eV wide) energy band centered on the atomic oxygen Kα fluorescence line at 0.53 keV. The X-ray power emitted from the rings in the 0.49–0.62 keV band is 84 MW, which is about one-third of that emitted from Saturn disk in the photon energy range 0.24–2.0 keV. Our analysis also finds a clear detection of X-ray emission from the rings in the 0.49–0.62 keV band in an earlier (14–15 April, 2003) Chandra ACIS observation of Saturn. Fluorescent scattering of solar X-rays from oxygen atoms in the $H_2O$ icy ring material is the likely source mechanism for ring X-rays, consistent with the scenario of solar photo-production of a tenuous ring oxygen atmosphere and ionosphere recently discovered by Cassini.

*Subject heading*: planets: rings — planets and satellites: individual (Saturn) — X-rays: general — scattering — Sun: X-rays — X-rays: individual (Saturn rings)

## 1. Introduction

The rings of Saturn, first seen in 1610 by Galileo Galilei, are one of the most fascinating objects in our solar system. The main ring system, from inside out, consists of the D (distance from Saturn, 1.11–1.235 $R_S$; Saturn radius $R_S$ = 60,3330 km), C (1.235–1.525 $R_S$), B (1.525–1.95 $R_S$), and A (2.025–2.27 $R_S$) rings (Cuzzi et al. 2002). These are followed by the fainter F, G, and E rings, which span 2.324–8.0 $R_S$. The rings are known to be made of mostly water ($H_2O$) ice (e.g., Esposito et al. 1984; Cuzzi et al. 2002). Recently, Cassini discovered a tenuous oxygen ionosphere, and therefore atmosphere, over the rings (Gurnett et al. 2005; Waite et al. 2005; Young et al. 2005). Here we report the discovery of oxygen Kα X-ray emission from the rings of Saturn. This result adds one more object to the list of solar system soft X-ray emitters found by the Chandra X-ray Observatory (CXO) during the last few years (e.g., Elsner et al. 2002; Dennerl 2002; Dennerl et al. 2002; Ness et al. 2004; see Bhardwaj et al. 2002 for a review of earlier studies).

## 2. Observations

Using the CXO's Advanced CCD Imaging Spectrometer (ACIS) we observed Saturn on 20 January, 2004, and 26–27 January, 2004. Each continuous observation lasted for about one full Saturn rotation (~37 ks, see Table 1 for observation details). The observations were carried out with the planetary image falling on the S3 CCD of the spectroscopy array in imaging mode, the configuration with the greatest sensitivity to X-ray energies below 1 keV. Pulse-height values of individual X-ray events were corrected for effects due to Saturn's optically bright disk in the same way as for Jupiter (Elsner et al. 2005); however, the corrections for Saturn are much smaller than for Jupiter. More details of the present observations are given in Bhardwaj et al. (2005a).

For each observation, the background rate was determined using a large region, free of cosmic sources, outside the planet and the rings. It is important to note, however, that the planet blocks all true X-rays from beyond Saturn's orbit. For the case of the rings, we note that there is no transmission of <3 keV X-ray photons through 1 mm of thick $H_2O$ ice, and 1 cm of thick ice can effectively block X-rays up to 10 keV (Henke et al. 1993). The main rings of Saturn are at least few meters thick (Salo & Karjalainen 2003) and the size of particles can range from centimeters to meters (French & Nicholson 2000). Therefore, any background contribution to the detected emission from the rings is mostly events due to charged particles, and the estimated background contributions quoted below are upper limits to the true background.

Chandra events are time tagged and can therefore be mapped into Saturn's rest frame using the online JPL HORIZONS ephemerides generator. The ring region itself was well defined using ellipses at the inner edge of the C ring and the outer edge of the A ring (see Fig. 1). In our analysis we excluded any portion of the rings overlapping the planet, whether in between the planet and the CXO as in the north or behind the planet as in the south. We note that Saturn's moons orbit in planes close to the ring plane, and no large (>~100 km) moons orbit within the main rings or at outskirts of the A ring. The orbits of large moons, such as Titan, did not place their projected sky images on the rings during these observations.

## 3. Results

Figure 1 shows the X-ray image of the Saturnian system on 20 and 26–27 January in the 0.49–0.62 keV band: the energy range where X-rays from the rings are unambiguously detected (see Fig. 2). The observations suggest that, similar to Saturn's X-ray emission (Bhardwaj et al. 2005a), the ring X-rays are highly variable — a factor of 2–3 variability in brightness over one week. Note the apparent asymmetry in X-ray emission from the east (morning) and west (evening) ansae of the rings on 20 January.

Figure 2 shows the background subtracted spectrum of ring X-rays for each of the two exposures. In the energy range 0.24–2.0 keV, where essentially the entire planet's X-ray emission is detected, the numbers of X-ray photons detected from the ring region defined above on 20 January and on 26–27 January are 65 and 23, respectively. Almost half of these photons, 28 and 14, have energies in the 0.49–0.62 keV band, and peak near the oxygen Kα fluorescence line emission at 0.53 keV. (Note that the energy resolution of the ACIS S3 CCD is ~120 eV at these energies). With the expected 0.49–0.62 keV background counts for the ring region being 3.0 and 1.5, respectively, the detection of ring X-rays in the 0.49–0.62 keV energy range is statistically highly significant for both exposures; the probability of detecting 28 (14) or more photons with only 3.0 (1.5) photons expected from the background is ~0 ($8 \times 10^{-10}$). A Gaussian with central energy of ~0.55 keV fits the observed ACIS spectrum quite well (see inset of Fig. 2), suggesting that the ring X-ray emission is due to O Kα emission. The X-ray power emitted by the rings in the 0.49–0.62 keV band on January 20 is 84 MW, which is about one-third of that emitted from the Saturn disk in the 0.24–2.0 keV band (Bhardwaj et al. 2005a).

Recently, Bhardwaj et al. (2005a) reported an X-ray flare from Saturn's disk in direct response to a solar X-ray flare on 20 January, 2004. They also showed that the temporal variation of the X-ray emission from Saturn's disk was similar to that of solar X-rays. In Figure 3(a) we plot the X-ray light curves for the rings on January 20 in 0.24–2.0 and 0.49–0.62 keV bands, as well as the expected background light curve in the 0.49–0.62 keV band. In Figure 3(b), we plot the X-ray light curves from the Sun as measured by GOES and TIMED/SEE satellites (at Earth) and from Saturn's disk (see Bhardwaj et al. 2005a for details). Although the X-ray light curve from rings shows some similarity to the disk light curve, no flaring is evident. The ring light curve has an average of 1.27 0.49–0.62 keV events per 30 minute bin. If a factor of 5 increase had occurred at the time of the flare from the planet's disk (Bhardwaj et al. 2005a), we would expect 6.35 photons in the corresponding time bin. Only 2 photons were detected. The probability of seeing 2 or less events expecting 6.35 is 0.05, corresponding to about a 2-σ deviation. On the other hand, the probability of seeing 2 or more events expecting 1.27 is 0.36, less than a 1-σ deviation. Thus, due to the low signal to noise ratio per bin for the ring light curve, we cannot confirm statistically the presence or absence of flaring from the rings. We note that the spectrum of solar X-rays during the flare is quite different (generally harder) than during the quiet period.

## 4. The 14–15 April, 2003, Chandra Observation

We re-analyzed the Chandra ACIS S3 Saturn observation of 14–15 April, 2003 (Ness et al. 2004) in the same manner as our January 2004 observations. Figure 4 shows the spectrum of ring X-rays on 14–15 April, 2003: a cluster of photons around the O-K$\alpha$ line is also evident in these CXO observations. Just as for our January 2004 CXO observations, the detection of ring X-ray emission in the 0.49–0.62 keV band is highly significant; the number of photons detected from the ring region in this band is 36 with only 6.5 expected from the background, the probability of chance occurrence of 36 or more events expecting 6.5 being $9\times10^{-16}$. The inset of Figure 4 shows the 0.49–0.62 keV image of the Saturnian system on 14–15 April. As pointed out by Ness et al. (2004), an excess of X-rays on east ansa of the rings is also seen in this image. During 14–15 April 2003 the X-ray power emitted by the rings in the 0.49–0.62 keV band is about 70 MW.

## 5. Discussion

The ring X-rays are unlikely to be produced by charged particle precipitation on the ring material because essentially no energetic particles are detected over the rings (Krimigis et al. 2005; Young et al. 2005). Particle precipitation can at most produce X-ray emission at the outer edge of the A-rings, but the CXO observations suggest X-ray emission largely from the B-ring. Also, no X-rays are expected from the plasma-atmosphere interaction over the rings, since Cassini observations indicate that the ring atmosphere is too thin (density $\sim 10^4$ to $10^5$ cm$^{-3}$; Waite et al. 2005).

The presence of O K$\alpha$ line emission suggests that the likely source mechanism of ring X-rays is fluorescent scattering of solar X-rays from oxygen atoms in the H$_2$O-icy rings. Taking the fluorescent yield of O K$\alpha$ as 0.0083 (Krause 1979), the average value of the 0–2.5 nm solar flux for 0–12 UT on 20 January from the TIMED/SEE measurements as $2\times10^{-4}$ W m$^{-2}$, the Sun-Saturn and Earth-Saturn distances from Table 1, and estimating the area of the rings from which we see X-ray emission to be about one-third that of Saturn's disk, we estimate an energy flux from the rings of about $5\times10^{-15}$ erg cm$^{-2}$ s$^{-1}$. (This area, from which we see ring X-ray emission, is smaller by a factor ~4–5 than the full area of the ring region that does not overlap the planet.) This value is similar to the observed energy flux derived from the XSPEC spectral fitting. This implies that fluorescent scatting of solar X-rays can power the X-ray emission from the rings of Saturn.

If solar X-ray radiation is the cause of the ring X-ray emission, as suggested by the simple calculation above, then the X-ray emission should have been uniformly distributed over the rings. However, the CXO observations suggest that the spatial distribution of ring X-ray emission is non-uniform (Figs. 1 and 4), with a concentration on east ansa (morning-side) of the rings. One possibility is a statistical fluctuation in the spatial distribution of ring X-rays due to the small number of observed photons. To test this, after appropriate scaling of coordinates, we combined the January 2004 and April 2003 CXO observations and divided the ring region into 12 sectors each 30° wide. Figure 5 shows the distribution of X-ray photons in the 0.49–0.62 keV band. We find 13 events from sector 1 and 25 events from diagonally opposite sector 7, both these sectors have equal projected area. The probability of detecting 25 or more photons expecting 13 is 0.002. If we combine 3 sectors on east and west ansae (sectors 6 to 8, and sectors 12, 1, and 2), the events are 37 and 27, respectively, and the probability in this case is 0.039. Thus, in the combined observational data, the evidence for non-uniform emission from the rings is suggestive but not overwhelmingly strong.

Spokes are an interesting feature of the Saturnian ring system, and have been observed over the rings, largely confined to the dense B-ring, and most often seen on the morning side (east ansa) (e.g., Cuzzi et al. 2002; Horányi et al. 2004; McGhee et al. 2005). Spoke lifetimes range from tens of minutes to a few hours. Spokes are clouds of fine ice-dust particles (~sub-micron to micron size) that are levitated off the ring surface, and suggested to be triggered by meteoritic impacts on the rings (Goertz & Morfill 1983; Cuzzi & Estrada 1998). Since the meteor impact is more likely in the midnight to early morning hours (due to larger relative velocities, like on Earth), the observation of spokes is more likely in the morning hours since the rings have recently emerged from Saturn's shadow (the nightside). The higher X-ray brightness on the morning side of the rings could be due to such meteoritic impacts exposing more ring ice for solar fluorescence, resulting in higher X-ray yield. Moreover, the icy-dust produced by the impact would also contribute to increased X-ray brightness in the morning sector by fluorescent scattering of solar X-rays, although the albedo for solar fluorescence from dust is expected to be relatively lower (Krasnopolsky 1997). Detailed modeling is required to calculate the solar fluorescence contribution from the icy-dust particles to ring X-rays.

Since scattering of solar x-rays takes place mostly in the top layer of the rings, the surface composition affects the X-ray scattering from the rings. In principle, X-ray observations could help determine the surface composition of the rings. However, the spatial resolution for the CXO at the distance of Saturn is a few 1000 km, and the X-ray flux from the rings is too small for realistic measurements at that distance. So, why are Saturn's main rings hard to see in X-rays, while they are so bright in visible light? Taking the mean visible albedo of the main rings as 0.5, a solar constant value of 1370 W m$^{-2}$ at Earth, and using the same area for the region of the rings from which we see X-ray emission as used to estimate the expected flux, the visible energy flux from the ring region at Earth is ~2 × 10$^{-6}$ erg cm$^{-2}$ s$^{-1}$. Since the energy of an X-ray photon from the rings is about 200 times that in the visible, the ratio of visible to X-ray photon flux is ~10$^{12}$: implying that only one X-ray photon is emitted from the rings for every 10$^{12}$ visible photons.

The present study suggests that Saturn rings shine in X-rays due to scattered solar radiation. The rings of Saturn now join the list of other solar system objects (like Mars, Venus, Moon, and the non-auroral disks of Jupiter, Saturn and Earth) that glow in soft X-rays via scattering of solar X-ray radiation (e.g., Dennerl 2002; Dennerl et al. 2002; Ness et al., 2004; Wargelin et al. 2004; Bhardwaj et al. 2005a, 2005b; cf. also Bhardwaj et al. 2002). The INMS and CAPS experiments on Cassini have recently discovered oxygen ions over the rings: suggesting a tenuous ring oxygen atmosphere likely produced by solar ultraviolet photon-induced decomposition of water-ice (Johnson et al. 2004; Waite et al. 2005; Young et al. 2005). Thus, recent Cassini and CXO observations suggest that solar UV–X-ray radiation plays an important role in physical and chemical processes in the rings of Saturn.

**Acknowledgements**

This research was performed while A. Bhardwaj held a National Research Council Senior Resident Research Associateship at the NASA Marshall Space Flight Center. This work is based on observations obtained with Chandra X-ray Observatory and was supported by a grant from the Chandra X-ray Center.

**Table 1.** Observational Parameters

| Parameter | Start time value[1] | Stop time value[1] |
|---|---|---|
| OBSID 4466 (4467) | 20, 00:05:02 (26, 14:30:24) | 20, 10:58:54 (27, 01:11:43) |
| R.A. (hh mm ss) | 06 35 26.75 (06 33 25.60) | 06 35 17.94 (06 33 17.98) |
| Dec (deg mm ss) | +22 33 17.4 (+22 35 51.6) | +22 33 28.6 (+22 36 01.8) |
| Sun Distance (AU[2]) | 9.034 (9.034) | 9.034 (9.034) |
| Earth Distance (AU) | 8.109 (8.155) | 8.112 (8.158) |
| Diameter[3] (arcsec) | 20.496 (20.380) | 20.489 (20.371) |
| Elongation[4] (degree) | 158.98/T[5] (151.79/T) | 158.48/T (151.31/T) |
| Phase[6] (degree) | 2.23 (2.95) | 2.28 (2.99) |

[1]2004 January: date, hh:mm:ss in UT. Values in parentheses are for OBSID 4467 on 26–27 January, 2004.
[2]AU = Astronomical Unit = $1.49598 \times 10^8$ km.
[3]Projected equatorial angular diameter of Saturn.
[4]Solar elongation = Sun-Earth-Saturn angle.
[5]T indicates Saturn trails Sun (evening sky).
[6]Phase = Sun-Saturn-Earth angle.

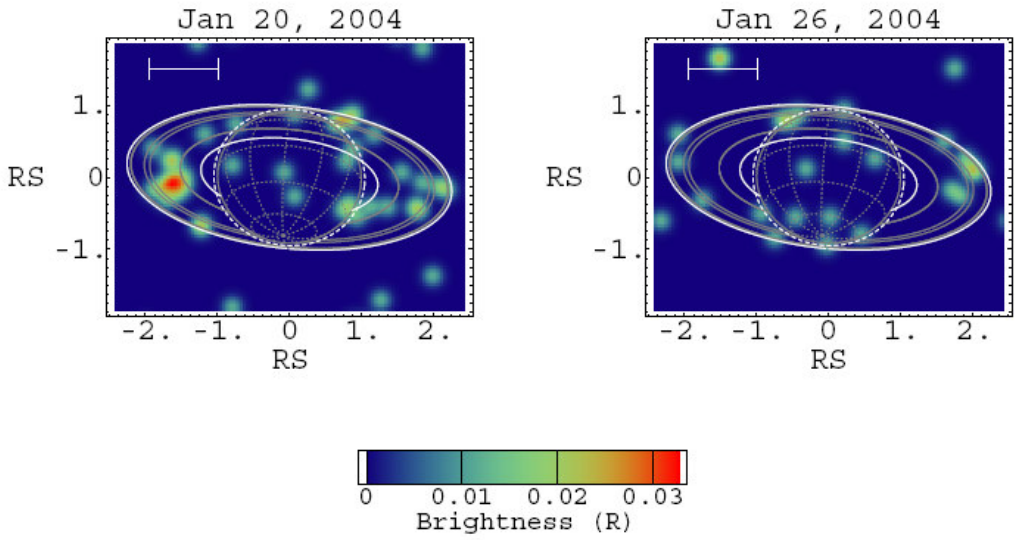

**Figure 1.** Chandra ACIS X-ray images of Saturnian system in the 0.49–0.62 keV band on 20 and 26-27 January, 2004. The X-ray emission from the rings is clearly present in these restricted energy band images (see Fig. 2); the emission from the planet is relatively weak in this band (see Fig. 1 of Bhardwaj et al. 2005a for an X-ray image of Saturnian system in the 0.24–2.0 keV band). The false color images, with brightness in Rayleighs (R), show X-ray photons as seen in a frame moving across the sky with Saturn, smoothed with a two dimensional gaussian with $\sigma = 0.984$ arcsecond (twice the ACIS pixel width). The conversion to Rayleighs used a value for effective area of 195 cm$^2$ at 0.525 keV — the energy of the atomic oxygen K$\alpha$ emission line. The horizontal and vertical axes are in units of Saturn's equatorial radius (RS). The white scale bar in the upper left of each panel represents 10 arcseconds. The superimposed graticule shows latitude and longitude lines at intervals of 30°. The solid grey lines are the outlines of the planet and rings, with the outer edge of A ring and inner edge of C ring shown in white. The dotted white line defines the region within which events were accepted as part of Saturn's disk unless obscured by the rings. The two images taken a week apart, and shown on the same color scale, indicate substantial variability in X-ray emission from the rings.

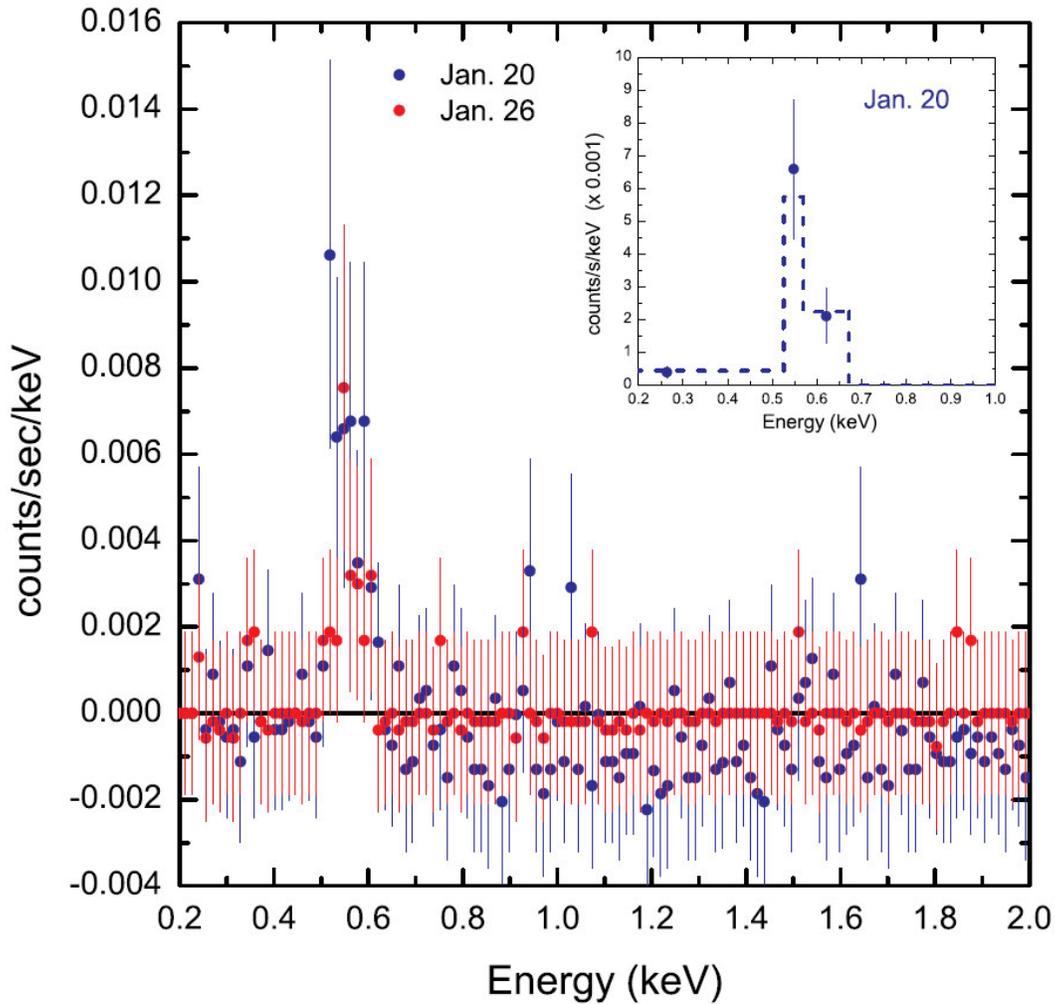

**Figure 2.** Background subtracted Chandra ACIS-S3 observed X-ray energy spectrum for Saturn's rings in the 0.2–2.0 keV range on 20 and 26-27 January, 2004. The cluster of X-ray photons in the ~0.49–0.62 keV band suggests the presence of the oxygen Kα line emission at 0.53 keV in the X-ray emission from the rings. The inset shows a Gaussian fit (peak energy = 0.55 keV, σ = 140 eV), shown by the dashed line, to the ACIS-observed rings spectrum on Jan. 20. Each spectral point (solid circle with error bar) represents ≥10 measured events. The spectral fitting suggests that X-ray emissions from the rings are predominantly oxygen Kα photons.

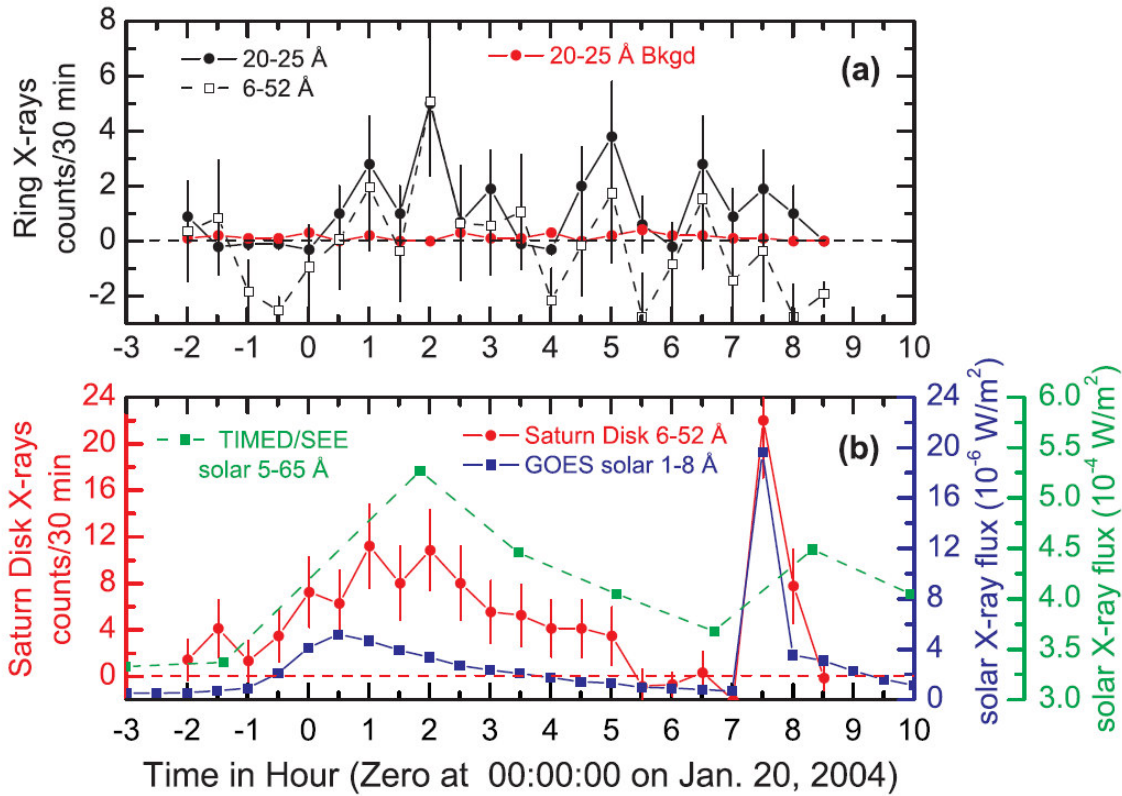

**Figure 3.** X-ray light curve for Saturn's rings and disk and the Sun on 20 January 2004. All data are binned in 30-minute increments, except for the TIMED/SEE data which are 3-minute observation-averaged fluxes obtained every orbit (~12 measurements per day). **(a)** Background-subtracted ring X-ray emission in 0.24–2.0 and 0.49–0.62 keV bands observed by Chandra-ACIS, plotted after shifting by -2.236 hr to account for the light travel time difference between Sun-Saturn-Earth and Sun-Earth. Expected background in the 0.49–0.62 keV band is shown in red. **(b)** Background-subtracted Saturn disk 0.24–2.0 keV emission (there are only 6 events in the 0.49–0.62 keV band from disk) plotted in red after shifting by -2.236 hr. The solar X-ray flux in the 1.6–12.4 keV band measured by the Earth-orbiting GOES-12 satellite is plotted in blue. The solar 0.2–2.5 keV fluxes measured by TIMED/SEE are solid green squares and are joined by the green dashed line for visualization purpose. A flare in the light curve for Saturn's disk and for the solar X-ray flux is observed at about 7.5 hr. Though the light curve for ring X-ray emission is somewhat similar to that for the disk X-ray emission, no flaring is evident.

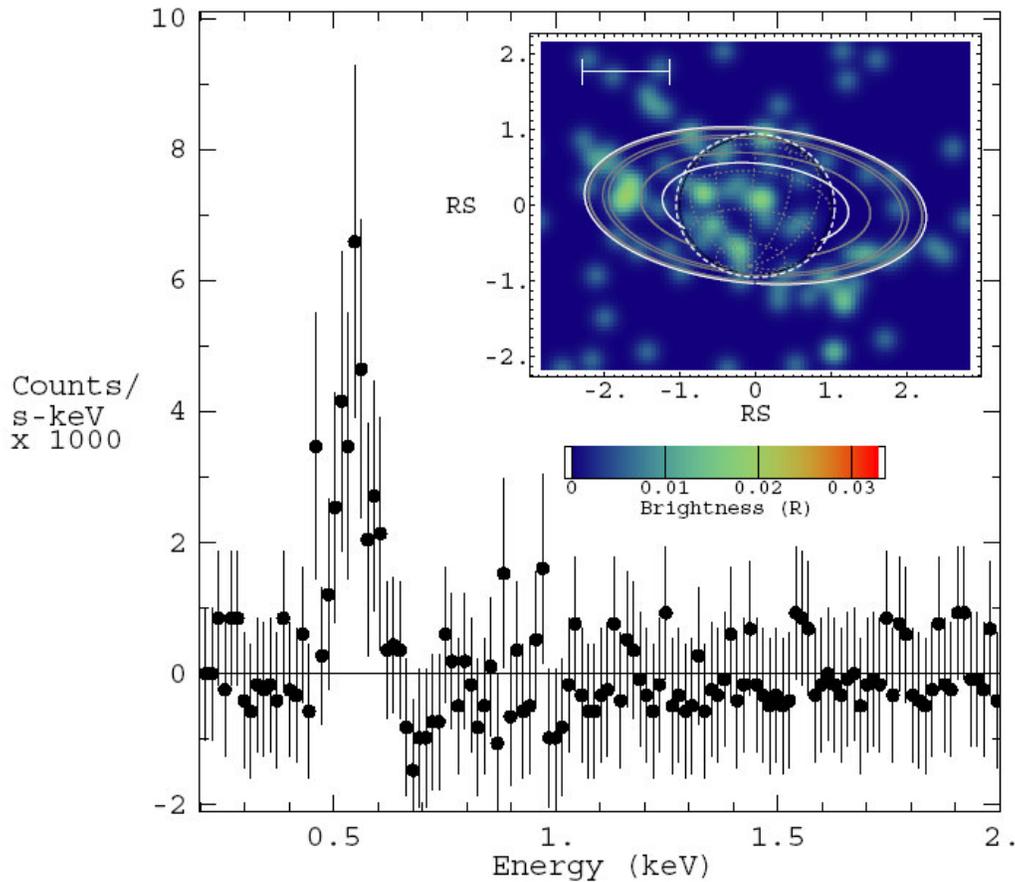

**Figure 4.** Background subtracted energy spectrum for ring X-ray emission during the Chandra ACIS 14–15 April 2003 observation (Ness et al. 2004). A cluster of X-ray photons in the ~0.49–0.62 keV band around the 0.53 keV oxygen Kα line is clearly evident in the spectrum. Note that the energy resolution of ACIS-S3 at these energies is ~120 eV. The inset shows the Chandra ACIS X-ray image of Saturn system in the 0.49–0.62 keV band on 14–15 April 2003. The description of this figure is same as for Fig. 1. Rings X-ray emission is evident on the east ansa (morning) in this image, which is also mentioned by Ness et al. (2004).

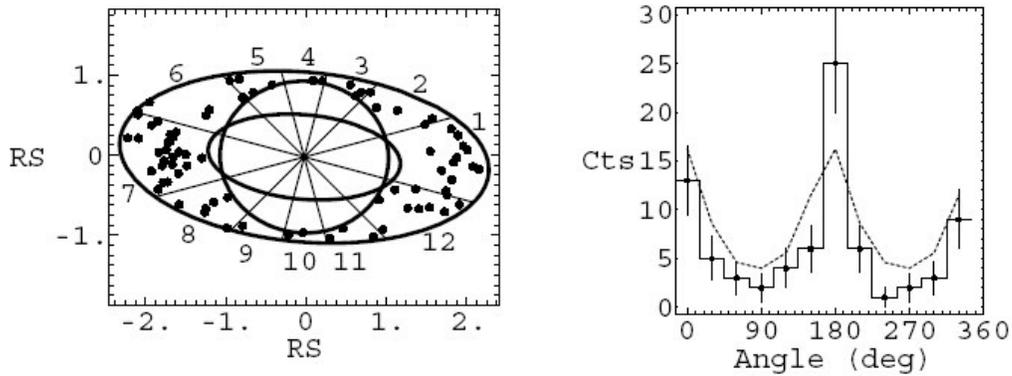

**Figure 5.** Distributions of X-ray events in the 0.49–0.62 keV band from the rings of Saturn obtained by combining the January 2004 and April 2003 Chandra ACIS observations. **Left:** Distribution in the ring plane in 30° wide sectors, with sector 1 ranging from -15° to +15° in azimuthal angle measured from the horizontal axis. Diagonally opposite sectors have equal area. **Right:** Distribution of counts in the 30° sectors of the left panel. The dotted curve shows the distribution expected if the emission from each sector were proportional to its area, normalized so that the total number of photons equals what is observed.